\documentclass{pasj01}
\Received{\today}
\Accepted{$\langle$acception date$\rangle$}
\Published{$\langle$publication date$\rangle$}
\SetRunningHead{Ozawa et al.}{JVLA S and X-band Polarimetry of the Merging Cluster Abell 2256}

\begin{document}

\title{JVLA S and X-band Polarimetry of the Merging Cluster Abell 2256}

\author{Takeaki \textsc{Ozawa}\altaffilmark{1}\thanks{Mail to: k5148776@kadai.jp}, Hiroyuki \textsc{Nakanishi}\altaffilmark{1}, Takuya \textsc{Akahori}\altaffilmark{1},\\ Kenta \textsc{Anraku}\altaffilmark{1}, Motokazu \textsc{Takizawa}\altaffilmark{2}, Ikumi \textsc{Takahashi}\altaffilmark{3},\\ Sachiko \textsc{Onodera}\altaffilmark{4}, Yuya \textsc{Tsuda}\altaffilmark{5}, and Yoshiaki \textsc{Sofue}\altaffilmark{6}}
\altaffiltext{1}{Graduate School of Science and Engineering, Kagoshima University, Korimoto 1-21-35, Kagoshima 890-0065, Japan}
\altaffiltext{2}{Department of Physics, Yamagata University, Kojirakawa-machi 1-4-12, Yamagata 990-8560, Japan}
\altaffiltext{3}{Graduate School of Science and Engineering, Yamagata University, Kojirakawa-machi 1-4-12, Yamagata 990-8560, Japan}
\altaffiltext{4}{Department of Physics, Meisei University, Hodokubo 2-1-1, Hino, Tokyo 191-8506, Japan}
\altaffiltext{5}{Graduate School of Science and Engineering, Hodokubo 2-1-1, Hino, Tokyo 191-8506, Japan}
\altaffiltext{6}{Institute of Astronomy, University of Tokyo, Osawa 2-21-1, Mitaka, Tokyo 181-0015, Japan}

\KeyWords{galaxies: clusters: individual: (Abell 2256) --- galaxies: clusters: intracluster medium --- magnetic fields --- polarization}

\maketitle

\begin{abstract}
We report polarimetry results of a merging cluster of galaxies Abell 2256 with Karl G. Jansky Very Large Array (JVLA). We performed new observations with JVLA at S-band (2051--3947~MHz) and X-band (8051--9947~MHz) in the C array configuration, and detected significant polarized emissions from the radio relic, Source A, and Source B in this cluster. We calculated the total magnetic field strengths toward the radio relic using revised equipartition formula, which is 1.8--5.0~$\mu \mathrm{G}$. With dispersions of Faraday rotation measure, magnetic-field strengths toward Sources A and B are estimated to be 0.63--1.26~$\mathrm{\mu G}$ and 0.11--0.21 $\mathrm{\mu G}$, respectively. An extremely high degree of linear polarization, as high as $\sim$ 35~\%, about a half of the maximum polarization, was detected toward the radio relic, which indicates highly ordered magnetic lines of force over the beam sizes ($\sim$ 52 kpc).The fractional polarization of the radio relic decreases from $\sim$ 35~\% to $\sim$ 20 \% around 3 GHz as the frequency decreases  and is nearly constant between 1.37 and 3 GHz. Both analyses with depolarization models and Faraday tomography suggest multiple depolarization components toward the radio relic and imply the existence of turbulent magnetic fields. 
 \end{abstract}

\section{Introduction}
\label{s1}

Collision of galaxy clusters is one of the most energetic events with kinetic energy on the order of $\sim~10^{64}$~ergs in the Universe \citep{2001ApJ...553L..15B}. Shock waves and turbulence induced by the collision can convert a huge kinetic energy of clusters into thermal/non-thermal energies of the intracluster medium (ICM). Cosmic-rays injected into the ICM by AGN activities, star formations of galaxies, and structure formation shocks \citep{2001MNRAS.320..365B} can be re-accelerated by the shocks \citep{2000ApJ...535..586T,2008Sci...320..909R,2009MNRAS.395.1333V} and turbulence \citep{2001MNRAS.320..365B, 2001ApJ...557..560P,2002ApJ...577..658O, 2003ApJ...584..190F, 2004MNRAS.350.1174B,2005MNRAS.357.1313C, 2009ApJ...698L..14X, 2010ApJ...725.2152X,2012A&ARv..20...54F,2013MNRAS.429.3564D}. The correlation between X-ray luminosity of the ICM and the power of diffuse radio emission from cosmic-rays is known for radio halos and relics (e.g., \cite{2012A&ARv..20...54F}). It suggests the relationship between the cluster size and the magnitude of  particle acceleration in sense that larger clusters can produce more powerful radio emission. However, the nature and evolution of the ICM and intergalactic magnetic field (IGMF), which determine the efficiency and the radio emission mechanisms, are poorly understood.

Turbulence is thought to play an important role in the evolution of the IGMF. It has been suggested that turbulence dynamo can amplify the IGMF in a cosmological time \citep{2008Sci...320..909R, 2009ApJ...705L..90C}. Actually, the Kolmogorov index in the power spectrum of magnetic fields has been reported (e.g. Abell 2362, \cite{2008A&A...483..699G}), indicating the existence of turbulence and amplification of the IGMF by turbulence in galaxy clusters.

One of the useful techniques to investigate turbulent magnetic fields is the depolarization, in which observed polarized intensity gets weaker than that at the origin arising from several mechanisms. Particularly, beam depolarizations of internal Faraday dispersion (IFD) and external Faraday dispersion (EFD) becomes significant effect, if structures of Faraday rotation measure (RM) inside and outside a polarized radio emission source, respectively, are not uniform within an observing beam, e.g. if small eddy-sized turbulent magnetic fields exist. Both IFD and EFD depend on the dispersion of RM within the beam. \citet{1966MNRAS.133...67B} analytically investigated the dependencies called Burn's law, and \citet{2011MNRAS.418.2336A} investigated the optimum frequency range for this technique.

Abell 2256 is known as a merging cluster of galaxies in which we expect turbulence of the ICM and turbulent magnetic fields in the cluster. In this paper, we report results of linear polarimetry of the central part of Abell 2256 with the Karl G. Jansky Very Large Array (JVLA) at S-band (2051--3947~MHz) and X-band (8051--9947~GHz) in the C array configuration. We obtained Stokes $I$, $Q$, and $U$ images in order to measure the total intensity, fractional polarization, and RM map. Using the obtained maps, we investigated IGMF structures in Abell 2256 by means of depolarization. The layout of this paper is as follows. In Section 2, we introduce Abell 2256. In Section 3, we describe the observations and data reductions in the JVLA. In Section 4, we present the results, which include the total intensity, fractional polarization and RM. In Section 5, we discuss the magnetic field strengths toward the radio relic, Source A, and Source B, and discuss the fractional polarization of the radio relic. In Section 6, we summarize our conclusions. 

Throughout this paper, we assume the following cosmological parameters: $H_0=70.5~\mathrm{km~s^{-1} ~ Mpc^{-1}}$, $\Omega_{\mathrm{m 0}}=0.27$, and $\Omega_{\mathrm{\Lambda 0}}=0.73$. The angular size of $\timeform{1'}$ corresponds to $\sim 67 ~ \mathrm{kpc}$ at the redshift of Abell 2256, $z=0.0581$, corresponding to a distance of $D=247$ Mpc.

\section{Cluster of galaxies Abell 2256}
\label{s2}

Abell 2256 is a nearby (redshift $z=0.0581$) cluster of galaxies whose X-ray center is located at (RA, Dec) = (\timeform{17h04m2.3s}, \timeform{+78D37'55.2''}) in the J2000 epoch \citep{1998MNRAS.301..881E}. \citet{2002AJ....123.2261B} investigated Abell 2256 with optical observations and found substructures of member galaxies with the peak radial velocity difference of $\sim 2000~ \mathrm{km~ s^{-1}}$. Substructures of the ICM are also known in X-ray observations (\cite{1991A&A...246L..10B, 1994Natur.372..439B, 2002ApJ...565..867S}). Using the X-ray satellite Suzaku, \citet{2011PASJ...63S1009T} estimated radial velocity difference of $\sim 1500 ~ \mathrm{km ~ s}^{-1}$ in gas bulk motions of the substructures. There are two distinct ICM components with temperatures $\sim 7~ \mathrm{keV}$ and $\sim 4.5~ \mathrm{keV}$ (\cite{2002ApJ...565..867S}). These results suggest that Abell 2256 is a merging galaxy cluster.

Radio observations have discovered a radio relic and a halo in the central part of Abell 2256 \citep{1976A&A....52..107B, 1979A&A....80..201B, 1994ApJ...436..654R, 2003AJ....125.2393M, 2006AJ....131.2900C, 2008A&A...489...69B, 2009A&A...508.1269V, 2010ApJ...718..939K, 2012A&A...543A..43V, 2014ApJ...794...24O, 2015A&A...575A..45T}. The radio relic located in the north-west of the cluster is $\sim 440$ ~kpc away from the X-ray center. The radio relic covers an area of $\timeform{16.9'} \times \timeform{7.8'} ~(1125~ \mathrm{kpc} \times 520~ \mathrm{kpc})$ \citep{2006AJ....131.2900C}. Previous observations revealed that the radio relic includes filamentary structures \citep{2006AJ....131.2900C, 2008A&A...489...69B, 2014ApJ...794...24O}. The radio halo is located in the central region of the cluster. \citet{2006AJ....131.2900C} measured that the total flux of the radio halo is approximately 103 mJy at 1369 MHz.

There are also several radio sources in Abell 2256. \citet{2003AJ....125.2393M} identified the radio sources associated with member galaxies of Abell 2256. Each radio source is labeled with alphabets \citep{1976A&A....52..107B, 1979A&A....80..201B, 1994ApJ...436..654R}, and Sources A, B, and C are the remarkable bright sources. Sources A and B are linearly polarized sources which are suitable for our measuring RMs. Source C is known as a head-tail galaxy which has a narrow straight tail extending for at least 480 kpc at 1.4 GHz across the radio relic \citep{1994ApJ...436..654R}. 

\section{Observations and Data reductions}
\label{s3}

\subsection{Radio observations}
\label{s3ss1}

\begin{table*}[tp]
\tbl{Details of the VLA \& JVLA observations of Abell 2256.}{%
\begin{tabular}{ccccccc}
\hline\hline
&Frequency\footnotemark[$*$] & Bandwidth\footnotemark[$*$] &  Configuration \footnotemark[$*$] & Date & Time\footnotemark[$*$] & Project\footnotemark[$*$] \\
&(MHz) & (MHz) & & & (h) & \\
\hline
VLA &1369/1417 & 25/25 & D & 1999-Apr-28 & 5.9, 5.9 & AC0522 \\
&1513/1703 & 12.5/25 & D & 1999-Apr-29 & 3.5, 5.5 & \\
\hline
VLA &1369/1417 & 25/25 & C & 2000-May-29 & 2.5, 2.5 & AC0545 \\
&1513/1703 & 12.5/12.5 & C & 2000-May-29 & 3.6, 3.6 & \\
&1369/1417 & 25/25 & C & 2000-Jun-18 & 2.5, 2.5 & \\
&1513/1703 & 12.5/25 & C & 2000-Jun-18 & 4.1, 3.5 & \\
\hline
JVLA &S-band 16 windows\footnotemark[$\dagger$] & 128 & C & 2013-Aug-25 & 1.2 & 13A-131 (this work)\\
& & & & 2013-Aug-26 & 1.2 & \\
& & & & 2013-Aug-29 & 1.2 & \\
 \hline
JVLA & X-band 16 windows\footnotemark[$\ddagger$] & 128 & C & 2013-Aug-18 & 1.3 & 13A-131 (this work)\\
& & & & 2013-Aug-19 & 1.3 & \\
\hline
\end{tabular}
}\label{t01}
\begin{tabnote}
\hangindent6pt\noindent
\hbox to6pt{\footnotemark[$*$]\hss}\unskip%
Column 2: observing frequency; Column 3: observing bandwidth; Column 4: array configuration; Column 5: dates of observation; Column 6: time on source; Column 7: NRAO project code.
\end{tabnote}
\begin{tabnote}
\hangindent6pt\noindent
\hbox to6pt{\footnotemark[$\dagger$]\hss}\unskip%
2051/2179/2307/2435/2563/2691/2819/2947/3051/3179/3307/3435/3563/3691/3819/3947.
\end{tabnote}
\begin{tabnote}
\hangindent6pt\noindent
\hbox to6pt{\footnotemark[$\ddagger$]\hss}\unskip%
8051/8179/8307/8435/8563/8691/8819/8947/9051/9179/9307/9435/9563/9691/9819/9947. 
\end{tabnote}
\end{table*}

We carried out new observations of Abell 2256 using JVLA at S-band (2051--3947 MHz) and X-band (8051--9947 MHz) in the C array configuration on 18--30 August 2013. Each band was separated into 16 spectral windows and each window had a bandwidth of 128 MHz (Table \ref{t01}). The pointing center was Source A ($\timeform{17h03m31s.9}, \timeform{+78D37'44''.4}$), since our observations primarily aimed at measuring RMs toward Sources A and B in the central part of Abell 2256. We observed 3C286 and 1803+784 as a flux and polarization calibrator, and a gain and phase calibrator, respectively. 

\subsection{Data reductions}
\label{s3ss2}
Data were reduced by the following procedures. Using National Radio Astronomy Observatory (NRAO) Common Astronomy Software Applications (CASA), we executed VLA calibration pipeline and task EXPORTUVFITS to convert JVLA's measurement sets into FITS so as to allow us to calibrate the data with NRAO Astronomical Image Processing System (AIPS). In AIPS, the data in each spectral window was averaged in frequency domain using task AVSPC. Radio Frequency Interferences (RFIs) and spurious signals were flagged using task CLIP and TVFLAG. We calibrated polarization using task PCAL and RLDIF after we created calibration tables. The data separated into observed date are concatenated using task DBCON. 

The data at 2179, 2307, 3691, 3819, and 3947 MHz in S-band are affected by RFIs. There were satellite downlink and digital audio radio service in 2180--2290 MHz and 3700--4200 MHz, respectively \footnote{https://science.nrao.edu/facilities/vla/docs/manuals/oss/performance/rfi}. Therefore, we removed the data in these spectral windows from our analysis.

In addition to the observed data, we also utilized the archival data which were observed with the VLA in the C and D array configurations at L-band (1369, 1417, 1512, 1703 MHz). We calibrated these data by the same procedures described above.

\begin{table*}[tp]
\tbl{Image qualities of total intensity and polarization at L, S, and X bands.}{%
\begin{tabular}{cccccc}
\hline\hline
& Frequency\footnotemark[$*$] & Beam\footnotemark[$*$] & $\sigma_{\mathrm{I}}$\footnotemark[$*$] & $\sigma_{\mathrm{Q}}$\footnotemark[$*$] & $\sigma_{\mathrm{U}}$\footnotemark[$*$] \\
& (MHz) & (\timeform{"}$\times$\timeform{"}) & ($\mathrm{mJy}~ \mathrm{beam}^{-1}$) & ($\mathrm{mJy}~ \mathrm{beam}^{-1}$) & ($\mathrm{mJy}~ \mathrm{beam}^{-1}$) \\
\hline
VLA & 1369 & 47$\times$47 & 0.163 & 0.028 & 0.022 \\
& 1417 & 47$\times$47 & 0.152 & 0.031 & 0.020 \\
& 1513 & 47$\times$47 & 0.183 & 0.041 & 0.028 \\
& 1703 & 47$\times$47 & 0.259 & 0.046 & 0.095 \\
JVLA & S-band 11 windows\footnotemark[$\dagger$]  & 47$\times$47 & 0.159 & 0.029 & 0.029 \\
& S-band 11 windows\footnotemark[$\dagger$]  & 15.1$\times$15.1 & 0.151 & 0.013 & 0.014 \\
& X-band 16 windows \footnotemark[$\ddagger$] & 15.1$\times$15.1 & 0.053 & 0.027 & 0.028 \\
\hline
\end{tabular}
}\label{t02}
\begin{tabnote}
\hangindent6pt\noindent
\hbox to6pt{\footnotemark[$*$]\hss}\unskip%
Column 2: observing frequency; Column 3: beam size; Columns 4,5,6: RMS noise of the Stokes $I$, $Q$, and $U$. We show the averaged RMS noise in JVLA S-band 11 windows and X-band 16 windows.
\end{tabnote}
\begin{tabnote}
\hangindent6pt\noindent
\hbox to6pt{\footnotemark[$\dagger$]\hss}\unskip%
2051/2435/2563/2691/2819/2947/3051/3179/3307/3435/3563.
\end{tabnote}
\begin{tabnote}
\hangindent6pt\noindent
\hbox to6pt{\footnotemark[$\ddagger$]\hss}\unskip%
8051/8179/8307/8435/8563/8691/8819/8947/9051/9179/9307/9435/9563/9691/9819/9947. 
\end{tabnote}
\end{table*}

We created Stoke $I$, $Q$ and $U$ images using task IMAGR with  suitable tapers. In order to detect the radio relic and resolve the individual polarized sources in the cluster, we made images of \timeform{47''} resolution for L and S bands, and \timeform{15.1''} resolution for S and X bands (Table \ref{t02}). Note that in the following analyses except Section 4.1, we used the images of \timeform{47''} and \timeform{15.1''} resolution for analyzing the radio relic and the individual radio sources, respectively. We excluded the data of X-band from the images of \timeform{47''} resolution, since the radio relic is outside the field of view. The images was convolved with a Gaussian beam using task CONVL. The number of pixels was $87 \times 87$ for images of \timeform{47''} resolution, and $136 \times 136$ for images of \timeform{15.1''}. Each pixel size corresponds to a half size of each beam size. 

\section{Results}
\label{s4}

\subsection{Radio images}
\label{s4ss1}
\begin{figure}[tp]
\begin{center}
\FigureFile(80mm,50mm){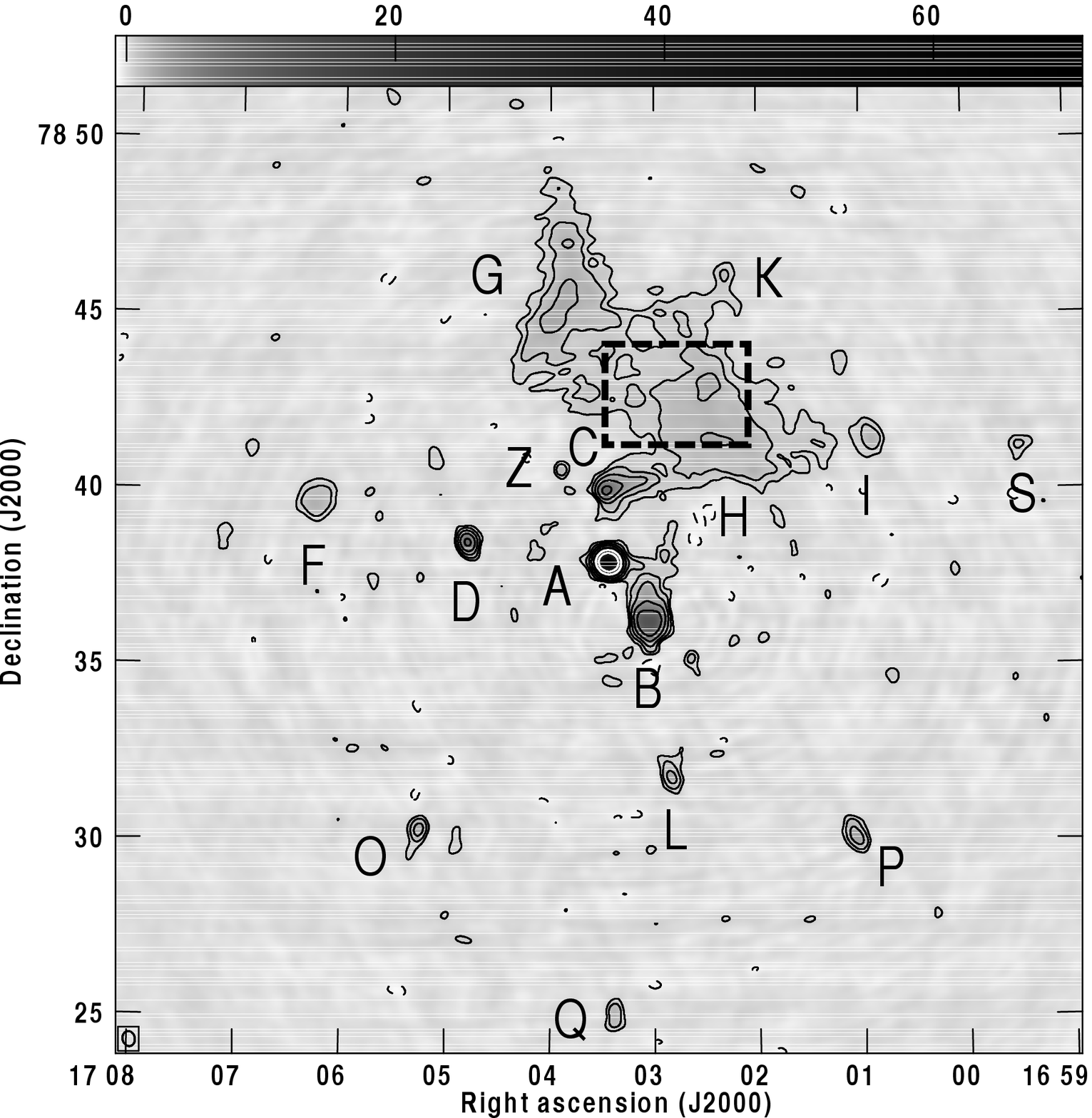}
\end{center}
\caption{
Total intensity map of Abell 2256 at 2051 MHz in the JVLA C array configuration. Contours are drawn at $(-3, 3, 6, 12, 24, 48, 96, 192) \times 140 ~ \mathrm{\mu Jy ~ beam^{-1}}$ $(9.702\times10^{-23}~\rm{W~m^{-2}~Hz^{-1}~sr^{-1}})$. The beam size of the image is shown at bottom-left and is \timeform{25.8''}$\times$\timeform{21.0''}. Radio sources are labeled following the references \citep{1976A&A....52..107B, 1979A&A....80..201B, 1994ApJ...436..654R}. The fractional polarization of the radio relic was measured in the dashed frame region (Section \ref{s4ss3}).}
\label{f01}
\end{figure}

Figure \ref{f01} shows the total intensity map of Abell 2256 at 2051 MHz. We detected several radio sources and the radio relic above 3$\sigma_\mathrm{I}$ significance. All of them are known sources in previous works with different array configurations and/or frequency bands (e.g. \cite{2003AJ....125.2393M, 2006AJ....131.2900C, 2014ApJ...794...24O}). In all of created images in S-band except the data affected by RFIs, we detected the radio relic, where an area of the relic above 3$\sigma_\mathrm{I}$ significance at 3563 MHz was smaller than that at 2051 MHz by a factor of $\sim 0.37$. 

\begin{figure}[tp]
\begin{center}
\FigureFile(80mm,50mm){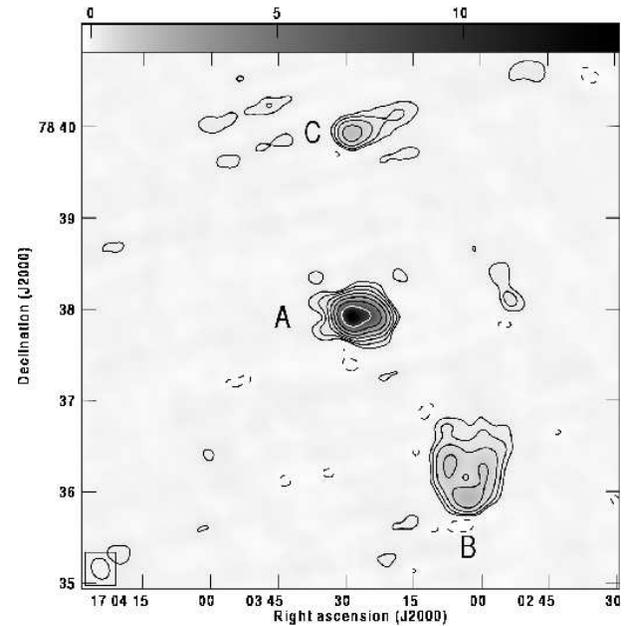}
\end{center}
\caption{Total intensity map of Sources A, B, and C at 8051 MHz in the JVLA C array configuration. Contour levels are drawn at $(-3, 3, 6, 12, 24, 48, 96, 192) \times 41.2 ~\mathrm{\mu Jy ~ beam^{-1}}$ $(9.424\times10^{-23}~\mathrm{W~m^{-2}~Hz^{-1}~sr^{-1}})$ . The beam size of the image is shown at bottom-left and is \timeform{14.4''}$\times$\timeform{11.4''}.}
\label{f03}
\end{figure}

Figure \ref{f03} shows the total intensity map at 8051 MHz. The radio relic is outside the field of view. In all of created images in X-band, we detected Sources A, B, and C. These radio sources are identified as radio galaxies (\cite{2003AJ....125.2393M}). 

\subsection{Total intensity}
\label{s4ss2}

The total 2051 MHz flux of the radio relic is estimated to be 286 mJy based on the image. For calculating the total 2051 MHz flux, we integrated the pixels where the flux density is above $3\sigma_I$ significance on the radio relic in the image of \timeform{47''} resolution. If we consider the total 1369 MHz flux from the radio relic of 462 mJy which is subtracted the flux of the point sources \citep{2006AJ....131.2900C}, and adopt the spectral index ($S \propto \nu^{\alpha}$) of $\alpha = -0.81$ \citep{2012A&A...543A..43V}, the total 2051 MHz flux should be 330 mJy. The measured total 2051 MHz flux of 286 mJy is thus smaller than the expected total flux by $\sim$ 13~\%. Note that the total 2051 MHz flux of 286 mJy is not subtracted the flux from the tail of Source C, so that the total flux of the relic is smaller than 286 mJy.

We also estimated the upper limit of the total 2051 MHz flux of the radio halo, which is $\sim49~\mathrm{mJy}$, assuming that the radius of the radio halo emission is $\sim\timeform{6.1'}$ \citep{2006AJ....131.2900C} and the upper limit of the flux density is $606~\mathrm{\mu Jy~beam^{-1}}~(10.296\times10^{-23}~\mathrm{W~m^{-2}~Hz^{-1}~sr^{-1}})$ which corresponds to 3$\sigma_\mathrm{I}$ significance at 2051 MHz. If we adopt the total 1369 MHz flux of the radio halo of 103 mJy \citep{2006AJ....131.2900C} and the spectral index of $\alpha = -1.1$ \citep{2012A&A...543A..43V}, the total 2051 MHz flux should be $\sim66$ mJy. Thus, the estimated upper limit of the total flux of 49 mJy is smaller than the expected total flux by $\sim26~\%$

\begin{figure}[tp]
\begin{center}
\FigureFile(80mm,50mm){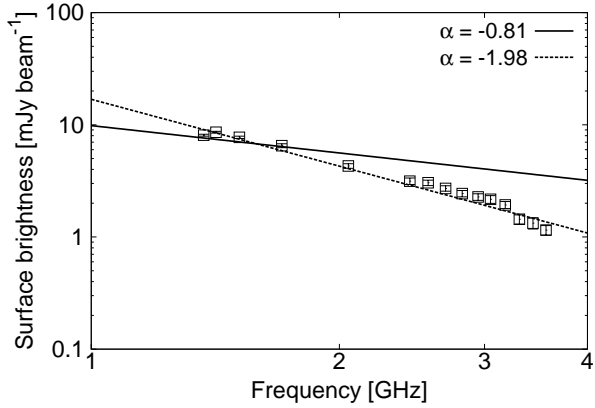}
\end{center}
\caption{An example of the spectral energy distribution (SED). Open squares show the SED for a pixel ($\timeform{17h02m51s.9}, \timeform{+78D42'26''.7}$) in the radio relic. The solid line represents an extrapolation of the spectrum from the surface brightness between 1369 MHz to 2051 MHz assuming the spectral index $\alpha = -0.81$. The dashed line represents a measured spectral index of $\alpha = -1.98$ from the surface brightness between 1369 MHz to 3563 MHz.
} 
\label{f04}
\end{figure}

Figure \ref{f04} shows an example of the spectral energy distribution (SED) at a point ($\timeform{17h02m51s.9}, \timeform{+78D42'26''.7}$) in the radio relic, where the SED is made from the images of \timeform{47''} resolution. We obtained a spectral index of $\alpha \sim -1.98$ from the surface brightness between 1369 MHz to 3563 MHz (the dashed line in Figure \ref{f04}). We found that the observed flux density of the radio relic at the frequency above 2 GHz is smaller than the extrapolation of the flux density from the results in 1369--2051 MHz with the spectral index $\alpha =-0.81$ (the solid line in Figure \ref{f04}).

A possible cause of such a decline is that we are missing the flux. In interferometry, we miss the flux from the diffuse emission which has a scale larger than the largest angular scale (LAS) of the interferometer. Actually, the scale of the major axis of the radio relic and halo is \timeform{1014''} and \timeform{732''} on the sky \citep{2006AJ....131.2900C}, while the LASs at 1.5 GHz and 3.0 GHz in the JVLA C array configuration were \timeform{970''} and \timeform{490''}, respectively. This possibility is also supported by single-dish radio observations using the Green Bank \citep{1975AJ.....80..263O} and Effelsberg \citep{1978A&AS...31...99H} telescopes, which yielded the total flux of 570 and 666 mJy in the entire area of Abell 2256 at 2695 MHz, respectively, in agreement with the spectral index of $\alpha =-0.72$ \citep{2008A&A...489...69B}.

Another possible cause of the decline would be a cutoff of cosmic-ray electrons at high energies. But since the effect of missing flux could be significant, we could not argue the possibility of the energy cutoff.

\subsection{Fractional polarization}
\label{s4ss3}
We detected significant polarized emission from the radio relic, Source A, and Source B at S-band. At X-band, we detected the significant polarized emission only from Source A. With the Stokes $I$, $Q$, and $U$, the fractional polarization $p$ is given by
\begin{equation}\label{s4ss3e1}
p=\frac{\sqrt{Q^2+U^2}}{I}.
\end{equation}
The fractional polarization are created from the images of \timeform{47''} and \timeform{15.1''} resolution to analyze the radio relic and the individual polarized sources. We calculated the fractional polarization only in the pixels where the flux densities of Stokes $I$, $Q$, and $U$ are all above $3\sigma$ significance. 

\begin{figure}[tp]
\begin{center}
\FigureFile(80mm,50mm){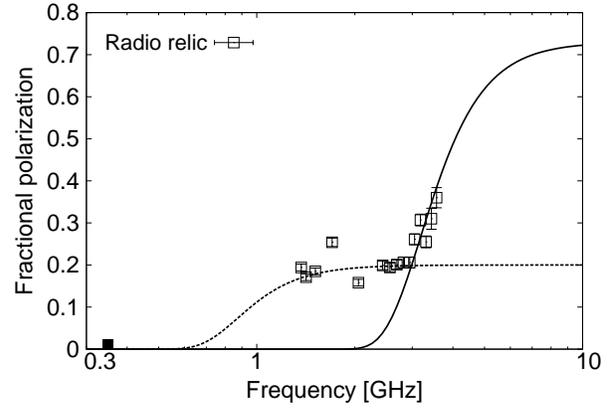}
\end{center}
\caption{The fractional polarization spectra of the radio relic (open squares). The filled square at the bottom-left is a observational result for the radio relic with the WSRT at 350~MHz \citep{2008A&A...489...69B}. The lines show the depolarization models defined in equation (\ref{s4ss3e2}) with $p_0=0.73$, $\sigma_{\mathrm{RM}}= 80~ \mathrm{rad~m^{-2}}$(solid line) and $p_0=0.2$, $\sigma_{\mathrm{RM}}= 6~ \mathrm{rad~m^{-2}}$(dashed line).}
\label{f05a}
\end{figure}

\begin{figure}[tp]
\begin{center}
\FigureFile(80mm,50mm){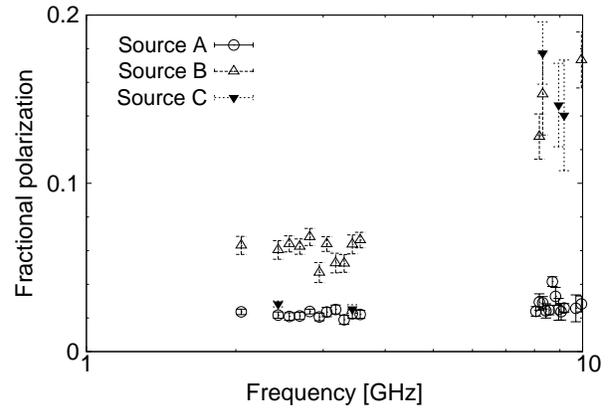}
\end{center}
\caption{The fractional polarization spectra of the Source A (open circles), Source B (open triangles), and Source C (filled inverted triangles).}
\label{f05b}
\end{figure}

Figure \ref{f05a} shows the fractional polarization spectra of the radio relic (open squares), and Figure \ref{f05b} shows the fractional polarization spectra of Source A (open circles), Source B (open triangles), and Source C (filled inverted triangles). Each data point represents a spatial average for the pixels within each emitting region. For the radio relic, we choose the region where the polarized emission is detected at 3563 MHz (the dashed frame region in Figure~\ref{f01}). We only plotted the data points which satisfy that the fractional polarization was obtained with at least 3 pixels within each emitting region in each frequency. The error bar indicates the standard deviation of the fractional polarization for the pixels. We found that the fractional polarization of the radio relic decreases from $\sim 35$~\% to $\sim 20$~\% as the frequency decreases from $\sim 3.5$~GHz to $\sim 3$~GHz. The fractional polarization is then nearly constant between 1.3--3~GHz. We plotted the fractional polarization of the brightest part of the radio relic observed at 350~MHz with Westerbork Synthesis Radio Telescope (WSRT) \citep{2008A&A...489...69B}. It indicates that the fractional polarization is less than 1~\%. Therefore, the fractional polarization in the field toward the relic varies twice, at $\sim 3$~GHz and around $\sim$ 0.4--1.3~GHz, and has step-like variations. 

We also see a change of the fractional polarization between 8051 MHz to 9947 MHz for Sources B and C. This is due to an artifact because sensitivity is not enough at X-band. For instance, the polarized intensity ($0.152~\mathrm{mJy}~\mathrm{beam}^{-1}$) is smaller than $3\sigma_{\mathrm{I}}$ significance ($0.159~\mathrm{mJy}~\mathrm{beam}^{-1}$) for a pixel (\timeform{17h03m07s8}, \timeform{+78D36'17''.8}) in the Source B, and the polarized intensity ($0.126~\mathrm{mJy}~\mathrm{beam}^{-1}$) is smaller than $3\sigma_{\mathrm{I}}$ significance ($0.145~\mathrm{mJy}~\mathrm{beam}^{-1}$) for a pixel (\timeform{17h03m28s1}, \timeform{+78D39'56''.2}) in the Source C at 8051MHz. We consider that weak polarized emission exist at X-band but we did not detect  the accurate polarized intensity due to low sensitivity.

\subsection{Faraday Rotation Measure} \label{s4ss4}

Faraday rotation measure (RM) is defined from the Faraday rotation of the linear polarization as
\begin{equation}\label{s4ss4e1}
\phi=\phi_0+\mathrm{RM}~\lambda^2,
\end{equation}
where $\phi$ is the observed polarization angle in radians, and $\phi_0$ is the intrinsic polarization angle in radians. RM is given by
\begin{equation}\label{s4ss4e2}
\mathrm{RM} \approx 812 \int n_eB_{\parallel}dl,
\end{equation}
in $\mathrm{rad} ~ \mathrm{m}^{-2}$, where $n_\mathrm{e}$ is the thermal electron density in $\mathrm{cm}^{-3}$, $B$ is the magnetic fields parallel to the line of sight in $\mu \mathrm{G}$, and $l$ is the path length in kpc. 

\begin{figure}[tp]
\begin{center}
\FigureFile(80mm,50mm){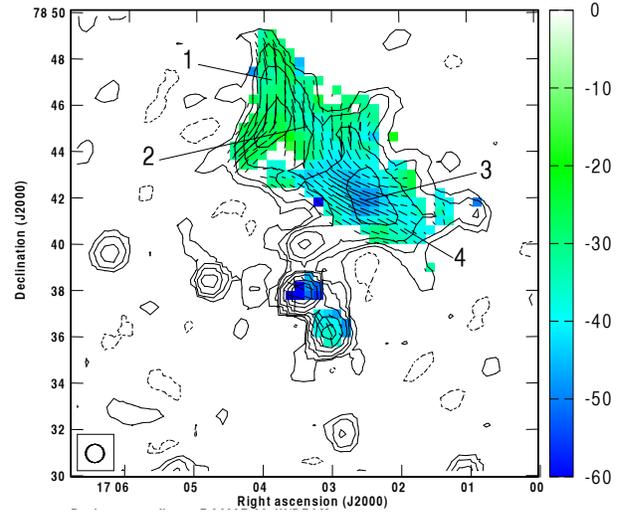}
\end{center}
\caption{The RM distribution of the Abell 2256 with the intrinsic B-vector. Also overlaid as contours is the total intensity map at 2051 MHz in the JVLA C array configuration drawn at $(-3, 3, 6, 12, 24, 48, 96, 192) \times 201.9 ~\mathrm{\mu Jy ~ beam^{-1}}$ $(3.432\times10^{-23}~\mathrm{W~m^{-2}~Hz^{-1}~sr^{-1}})$, convolved with a circular Gaussian beam with a FWHM of \timeform{47''}. The numbers represent the positions chosen in Figure~\ref{f07}.}
\label{f06}
\end{figure}

\begin{figure}[tp]
\begin{center}
\FigureFile(80mm,50mm){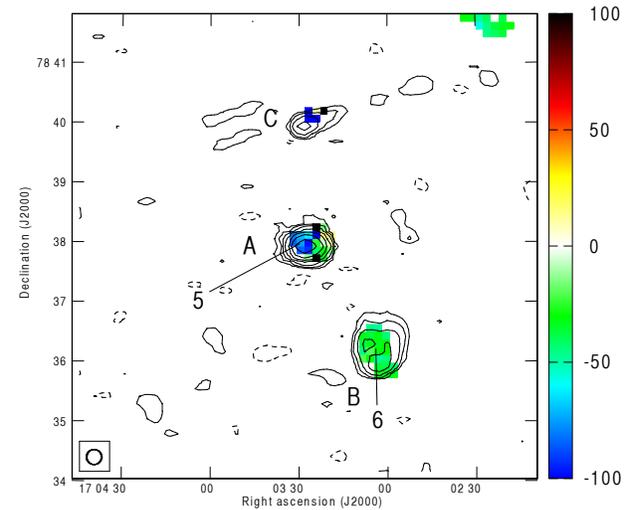}
\end{center}
\caption{The RM distribution of the Sources A, B, and C. Also overlaid as contours is the total intensity map at 8051 MHz in the JVLA C array configuration drawn at $(-3, 3, 6, 12, 24, 48, 96, 192) \times 48.4 ~\mathrm{\mu Jy ~ beam^{-1}}$ $(7.976\times10^{-23}~\mathrm{W~m^{-2}~Hz^{-1}~sr^{-1}})$, convolved with a circular Gaussian beam with a FWHM of \timeform{15.1''}.}
\label{f12}
\end{figure}

The RM map was created according to the linear relation between $\phi$ and $\lambda^2$ in equation~(\ref{s4ss4e1}). We used the polarization angle images of \timeform{47''} and \timeform{15.1''} resolution to analyze the radio relic and the individual polarized sources. We calculated RM only in the pixels which satisfy the following conditions: the flux density of the Stokes $I$, $Q$, and $U$ are all above $3\sigma$ significances, and the pixels satisfying the first condition are available from at least 4 frequencies. The RM map of Abell 2256 is shown in Figure \ref{f06} and \ref{f12}. 

\begin{figure}[tp]
\begin{center}
\FigureFile(80mm,50mm){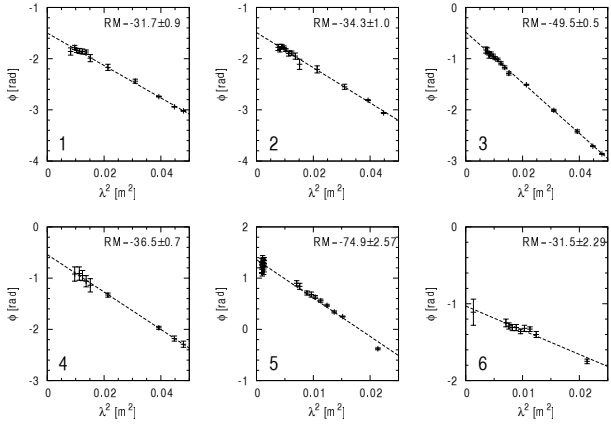}
\end{center}
\caption{Sample plots of the polarization angle $\phi$ against $\lambda^2$ for different positions in Abell 2256. Each position is shown in Figure~\ref{f06} and \ref{f12}.}
\label{f07}
\end{figure}

To make sure that our RMs based on a linear-fit between $\phi$ and $\lambda^2$ are reasonable, we examined the $\phi$--$\lambda^2$ relations toward the radio relic, Source A, and Source B. Figure~\ref{f07} shows some examples for the positions inside them (see Figure~\ref{f06} and \ref{f12}). We confirmed that the linear relation is roughly satisfied for the radio relic, Source A, and Source B. We also detected the RMs from the Source C but we did not use these RMs, since the polarized emission from Source C is not significant at X-band due to low sensitivity (see Section \ref{s4ss3}).

\begin{table}[tp]
\tbl{The average and standard deviation of RM.}{%
\begin{tabular}{cccc}
\hline\hline
Target & $\langle \mathrm{RM} \rangle$\footnotemark[$*$]
 & $\sigma_{\mathrm{RM}}$\footnotemark[$*$] & reference\\
       & ${\rm rad~m^{-2}}$             & ${\rm rad~m^{-2}}$\\
\hline
Relic    &   -44 &    7 & \citet{2006AJ....131.2900C}\\
Relic    & -34.5 &  6.2 & this work \\
Source A & -24.9 & 65.5 & this work \\
Source B & -34.1 & 10.5 & this work \\
\hline
\end{tabular}
}\label{t03}
\begin{tabnote}
\hangindent6pt\noindent
\hbox to6pt{\footnotemark[$*$]\hss}\unskip%
$\langle \mathrm{RM} \rangle$ and $\sigma_{\mathrm{RM}}$ are the average and standard deviation of RM, respectively.
\end{tabnote}
\end{table}

We calculated the average, $\langle \mathrm{RM} \rangle$, and the standard deviation, $\sigma_{\mathrm{RM}}$, of RM for the radio relic, Source A, and Source B. The results are listed in Table~\ref{t03}. We also show the results for the radio relic reported by \citet{2006AJ....131.2900C}. Our results for the radio relic and Source B are broadly consistent with the previous estimations for the radio relic. On the other hand, Source A has substantially smaller $\langle \mathrm{RM} \rangle$ and much larger $\sigma_{\mathrm{RM}}$ compared to the other positions. \citet{2004A&A...424..429M} reported that the simulated $|\langle \mathrm{RM}\rangle|/\sigma_\mathrm{RM}$ ratio depends only on the magnetic field power spectrum slope, and it has a considerable scatter. We consider that the smaller $\langle \mathrm{RM} \rangle$ and larger $\sigma_{\mathrm{RM}}$ could be related to the magnetic field fluctuations in the cluster.

\section{Discussion}
\label{s5}

\subsection{Magnetic Field Strengths in the Radio Relic}
\label{s5ss2}
We calculated the magnetic field strengths of the radio relic in Abell 2256 using the revised equipartition formula from \citet{2005AN....326..414B}. The total equipartition magnetic field strengths $B_\mathrm{t}$ is given by
\begin{equation} \label{s5ss2e1}
B_\mathrm{t} = \Biggl\{ \frac{4\pi(1-2\alpha)(K_{0}+1) I_{\nu} E_\mathrm{p}^{1-2\alpha} (\nu/2c_{1})^{-\alpha}}{(-2\alpha-1)c_2 (\alpha)lc_4(i)} \Biggr\}^{1/(3-\alpha)},
\end{equation}
where $B_\mathrm{t}$ is in G, $\alpha$ is the synchrotron spectral index\footnote{We use the definition of $S \propto \nu^{\alpha}$.}, $K_0$ is the ratio of the number densities of protons and electrons, $I_\nu$ is the synchrotron intensity at frequency $\nu$, $E_\mathrm{p}$ is the proton rest energy, and $l$ is the path length through the radio relic. The constants $c_1$, $c_2$, and $c_3$ are defined as
\begin{eqnarray*}
c_1 &=& \frac{3e}{4\pi m_\mathrm{e}^3 c^5}=6.26428 \times 10^{18}~ \mathrm{erg^{-2}~s^{-1}~G^{-1}} \\
c_2 &=& \frac{1}{4}c_3 \frac{(\gamma + 7/3)}{(\gamma +1)} \Gamma[(3\gamma -1)/12]\times \Gamma[(3\gamma +7)/12] \\
c_3 &=& \frac{\sqrt{3}e^3}{(4\pi m_\mathrm{e} c^2)}=1.86558\times10^{-23}~ \mathrm{erg~G^{-1}~sr^{-1}} \\
c_4 &=& [\cos (i)]^{(\gamma+1)/2},
\end{eqnarray*}
where $e$ is the elementary charge, $m_\mathrm{e}$ is the electron mass, $c$ is the speed of light, $\gamma$ is the spectral index of the electron energy spectrum which relates to the synchrotron spectral index $\alpha=-(\gamma-1)/2$, and $i$ is the inclination of the magnetic fields with respect to the sky plane \citep{2005AN....326..414B}. 

We obtained averaged synchrotron intensity $I_\nu$ of $6.27\times10^{-19}~\mathrm{erg~s^{-1}~cm^{-2}~Hz^{-1}~sr^{-1}}$ at 2051 MHz in the radio relic where the fractional polarization was measured (the dashed frame region in Figure \ref{f01}). On the other hand, we adopted the spectral index of $\alpha = -0.81$ measured by \citet{2012A&A...543A..43V} since we could not measure the accurate spectral index from our JVLA data due to the missing flux (see Section \ref{s4ss2}). We assumed the ratio of the proton--electron number densities of $K_0=100$, which is consistent with the acceleration process of cosmic-ray electrons for secondary particles and turbulence. Since the radio relic has a 25 kpc thickness at the minimum \citep{2014ApJ...794...24O} and covers $\sim1125~\mathrm{kpc}\times520~\mathrm{kpc}$ \citep{2006AJ....131.2900C}, we assumed the path length $l$ of $25$ kpc and $1125$ kpc. For the inclination, we assumed a mid-value of $i=\timeform{45D}$. 

We obtained the total equipartition magnetic field strengths of the radio relic of $\sim5.0~\mathrm{\mu G}$ with $l=25~\mathrm{kpc}$ and $\sim1.8~\mathrm{\mu G}$ with $l=1125~\mathrm{kpc}$. These values of micro-Gauss order  are consistent with the values of $1.5^{+0.9}_{-0.6}~\mathrm{\mu G}$ and $3.3^{+2.0}_{-1.2}~\mathrm{\mu G}$ with $\alpha=-1.25$ using the classical and the hadronic minimum energy conditions,  estimated by \citet{2006AJ....131.2900C}, respectively. 

To obtain the uniform and random magnetic field strengths, we can use the relationship between the observed fractional polarization $p$ and the degree of uniformity $f$ of the magnetic fields \citep{1976Natur.264..222S}
\begin{equation} \label{s5ss2e2}
p=\frac{3\gamma+3}{3\gamma+7} \Biggl[ 1+ \frac{(1-f)\pi^{1/2} \Gamma[(\gamma+5)/4]}{2f(\sin \theta)^{(\gamma+1)/2}\Gamma[(\gamma+7)/4]} \Biggr]^{-1},
\end{equation}
where $\Gamma$ is the gamma function and $\theta$ is the angle between the line of sight and the uniform magnetic fields. The ratio between the strengths of the uniform magnetic fields $B_\mathrm{u}$ and total magnetic fields $B_\mathrm{t}$ is given by $B_\mathrm{u}/B_\mathrm{t}=f^{2/(\gamma+1)}$ \citep{1982A&A...106..121B}. The random magnetic field strengths $B_\mathrm{r}$ is $B_\mathrm{r}=(B_\mathrm{t}^2 - B_\mathrm{u}^2)^{1/2}$.

 In order to avoid the effect of the depolarization, we used the fractional polarization of high frequency, which is $p=0.36$ at $3563~\mathrm{MHz}$. We also assumed the mid-value of $\theta=\timeform{45D}$.

We obtained the degree of uniformity of $f\sim 0.56$, which indicates that there are uniform magnetic fields in the radio relic with random magnetic fields. The uniform magnetic field strengths is $\sim3.7~\mathrm{\mu G}$ with $l=25~\mathrm{kpc}$ and $\sim1.3~\mathrm{\mu G}$ with $l=1125~\mathrm{kpc}$, and the random magnetic field strengths is $\sim3.4~\mathrm{\mu G}$ with $l=25~\mathrm{kpc}$ and $\sim1.2~\mathrm{\mu G}$ with $l=1125~\mathrm{kpc}$. However, $f\sim 0.56$ could be more large value because equation (\ref{s5ss2e2}) does not take into account the depolarization. We can see the ordered intrinsic magnetic fields over the beam size of $\sim52$~kpc in Figure \ref{f06} against $f\sim 0.56$. This could indicate that there are random magnetic fields along the line of sight toward the radio relic and depolarization occurred. 

\subsection{Magnetic Field Strengths in the Intracluster Space}
\label{s5ss3}

We estimated magnetic-field strengths in Abell 2256 using a traditional cell model \citep{1982ApJ...252...81L, 1991MNRAS.250..726T}. In the model, we consider cells along the line of sight from the observer to the polarized source, and each cell consists of uniform size, uniform electron density, and uniform magnetic field strength with a single scale and random field orientation. In this case, distribution of RM becomes the Gaussian with zero mean, and the variance of RM is given by
\begin{equation}\label{s5ss3e1}
\sigma_{\mathrm{RM}}^2=\frac{812^2\Lambda_\mathrm{B}}{3} \int(n_\mathrm{e}B_{\parallel})^2dl,
\end{equation}
where $\Lambda_\mathrm{B}$ is the cell size in kpc. For the distribution of thermal electron density, we adopt the $\beta$-model:
\begin{equation}\label{s5ss3e2}
n_\mathrm{e}=n_0(1+\frac{r^2}{r_\mathrm{c}^2})^{-3\beta /2},
\end{equation}
where $n_0$ is the central electron density in $\mathrm{cm}^{-3}$, $r$ is the distance from the X-ray center in kpc, and $r_\mathrm{c}$ is the core radius in kpc. Adopting equation~(\ref{s5ss3e2}) into equation~(\ref{s5ss3e1}), we obtain
\begin{equation}\label{s5ss3e3}
\sigma_{\mathrm{RM}}=\frac{KBn_0r_\mathrm{c}^{1/2} \Lambda_\mathrm{B}^{1/2}}{(1+r^2/r_\mathrm{c}^2)^{(6\beta -1)/4}} \sqrt{\frac{\Gamma(3\beta -0.5)}{\Gamma(3\beta)}},
\end{equation}
where $B=\sqrt{3B_{\parallel}}$ considering isotropic fields and $\Gamma$ is the Gamma function. $K$ is the constant which depends on the position of a backside polarized source along the line of sight; $K=624$ if the source is located behind the cluster and $K=441$ if the source is located at a halfway of the cluster \citep{1995A&A...302..680F, 1996ASPCP...88...271, 2010A&A...522A.105G}. Therefore, with $n_0$, $r$, $r_\mathrm{c}$, $\beta$ and $\Lambda_B$, equation (\ref{s5ss3e3}) leads the magnetic field strength along the line of sight.

\begin{table*}[tp]
\tbl{Parameters for magnetic field strengths in Abell 2256.}{%
\begin{tabular}{llccccccc}
\hline\hline
Source & $K$ &$\sigma_{\mathrm{RM}}$ & $n_0$\footnotemark[$*$] & $r$\footnotemark[$\dagger$] & $r_\mathrm{c}$\footnotemark[$*$] & $\beta$\footnotemark[$*$] & $\Lambda_\mathrm{B}$ & $B$  \\
 & & $\mathrm{[rad ~m^{-2}]}$ & $\mathrm{[10^{-3}~cm^{-3}]}$ & $[\mathrm{kpc}]$ & $[\mathrm{kpc}]$ & & $[\mathrm{kpc}]$ & [$\mu \mathrm{G}$] \\
\hline
Abell 2256 A & 441&65.5  & 2.6 & 7.2     & 587 & 0.914 & 20--5 & 0.63--1.26\\
Abell 2256 B & 441&10.5  & 2.6 & 133.7 & 587 & 0.914 & 20--5 & 0.11--0.21\\
\hline
\end{tabular}
}\label{t05}
\begin{tabnote}
\hangindent6pt\noindent
\hbox to6pt{\footnotemark[$*$]\hss}\unskip%
Reference: \footnotemark[$*$]\citet{2007A&A...466..805C} ; \footnotemark[$\dagger$]\citet{1998MNRAS.301..881E}.
\end{tabnote}
\end{table*}

The adopted parameters and results are shown in Table~\ref{t05}. Here, we consider $\Lambda_B$ from 5 to 20 kpc according to a dynamo theory \citep{2009ApJ...705L..90C}. We obtained that the field strength toward Source A is $B=$ 1.26 $\mu \mathrm{G}$ with $\Lambda_B=5$ kpc and $B=$ 0.63 $\mu \mathrm{G}$ with $\Lambda_B=20$ kpc, and the field strength toward Source B is $B=$ 0.21 $\mu \mathrm{G}$ with $\Lambda_B=5$ kpc and $B=$ 0.11 $\mu \mathrm{G}$ with $\Lambda_B=20$ kpc. 

\subsection{Contribution of the Galactic Magnetic Fields}
\label{s5ss4}

\begin{figure}[tp]
\begin{center}
\FigureFile(80mm,50mm){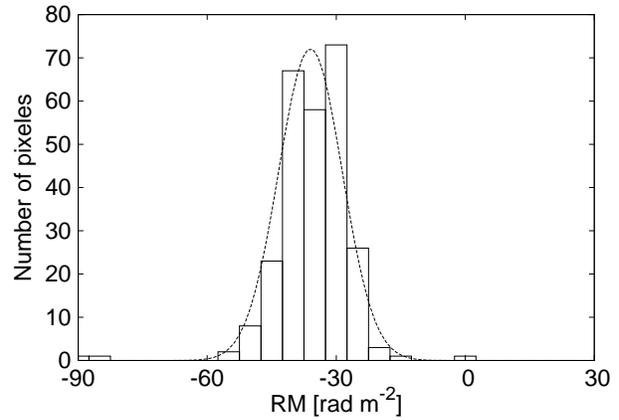}
\end{center}
\caption{Histogram of the RM. We obtained 355 pixels in the RM map of Abell 2256. Dashed line represents the result of Gaussian fitting with $\mu=-36~\mathrm{rad~m^{-2}}$ and $\sigma=7.4~\mathrm{rad~m^{-2}}$.}
\label{f08}
\end{figure}

Table~\ref{t03} suggests a shift of the mean of the RM from $0~ \mathrm{rad ~ m}^{-2}$ to about $-30~\mathrm{rad ~ m}^{-2}$ toward the Abell 2256 field. Figure~\ref{f08} shows the histogram of RMs in the Abell 2256 field. We obtained 355 pixels in Figure~\ref{f06}, and they actually indicates a histogram centered around $-36~\mathrm{rad ~ m}^{-2}$.

We consider that the shift is due to the Galactic contribution. In order to estimate the Galactic contribution to the Abell 2256 field, we examined the RM values of 28 polarized sources within 6\timeform{D} around Abell 2256 using the all-sky RM catalogue \citep{tss09}, and calculated the average of their RMs. We found that the average of the RMs for 28 polarized sources is $-30.0~ \mathrm{rad ~ m}^{-2}$ with the standard deviation of $16.7~ \mathrm{rad ~ m}^{-2}$. The average is broadly consistent with the means of RM for the radio relic, Source A, and Source B.

\subsection{Step-like Variations of the Fractional Polarization}
\label{s5ss5}

The fractional polarization of the radio relic varies at $\sim$ 3~GHz and around $\sim$ 0.4--1.3~GHz, and gives step-like variations as shown in Figure \ref{f05a}. Such a decrease of the fractional polarization toward low frequencies implies that depolarization takes place. If this is the case, we could investigate turbulent magnetic fields along the line of sight, as introduced in Section~\ref{s1}. We can analytically model the fractional polarization in the cases of the EFD and IFD using the Burn's law \citep{1966MNRAS.133...67B}. The fractional polarizations of EFD and IFD can be written as 
\begin{equation}\label{s4ss3e2}
p_\mathrm{{EFD}}=p_0 e^{-S},
\end{equation}
and
\begin{equation}\label{s4ss3e3}
p_\mathrm{{IFD}}=p_0\frac{1-e^{-S}}{S},
\end{equation}
respectively, where $p_0$ is the intrinsic fractional polarization, $S=2\sigma_{\mathrm{RM}}^2 \lambda^4$, $\sigma_{\mathrm{RM}}$ is the standard deviation of RM within the field of consideration, and $\lambda$ is the wavelength. The Burn's law is, however, a function which does not produce a step-like variation of the fractional polarization. Figure~\ref{f05a} shows an example of the EFD, and clearly indicates that a single depolarization component is not enough to explain the observed step-like variations of the radio relic. 

In addition to the depolarization, we suspect that the missing flux also affects the fractional polarization above $\sim 3$~GHz. If there is large diffuse source larger than the LAS, we only detect the compact diffuse source smaller than the LAS. If the fractional polarization of the compact diffuse source is larger than that of the large diffuse source, then the fractional polarization could increase as the observing frequency increases. Therefore, the variation of the fractional polarization at 3--3.5~GHz may be partly due to the missing flux.

Since the fractional polarizations of Sources A and B, which are located near the radio relic, does not show the variation at $\sim 3$~GHz, its origin should be significant for the emission from the radio relic. Indeed, because Sources A and B are compact sources, the effect of the missing flux is expected to be insignificant for the emissions from Sources A and B. Thus, to clarify the effect of the missing flux on the fractional polarization, single dish observations at 3--3.5~GHz should be performed in future.

On the other hand, the variation of the fractional polarization expected at around $\sim 1$~GHz could not be explained by the missing flux, since the LAS at $\sim 1$ GHz or less is sufficiently larger than the the major axis of the radio relic.

\subsection{Depolarization toward the Radio Relic}
\label{s5ss6}
\begin{table*}[tp]
\tbl{Parameters for the Depolarization Models.}{%
\begin{tabular}{ccccccccc}
\hline\hline
Model & Component & $B$ & $n_\mathrm{e}$ & $\Delta l$ & $N_\mathrm{X} \times N_\mathrm{Y}$  & $N_\mathrm{Z}$ & Intensity & $\sigma_{\mathrm{RM}}$ \\
& & $[\mu \mathrm{G}]$ & $\mathrm{[10^{-3}~cm^{-3}]}$ & $[\mathrm{kpc}]$ & $[\mathrm{kpc} \times \mathrm{kpc}]$ & $[\mathrm{kpc}]$ & & $\mathrm{[rad ~m^{-2}]}$ \\
\hline
EFD+EFD & foreside & 0.2 & 1.0 & 1\footnotemark[$*$]  & $50\times50$ \footnotemark[$*$]  & 600 & 1 & 3.2\\
 & backside& 2.3 & 3.0 &  & & 600 & 3 & 111\\
IFD+IFD  & foreside& 0.2 & 1.0 & 5 \footnotemark[$*$]  & $50\times50$ \footnotemark[$*$]  & 500 & 1 & 4.1\\
& backside& 2.3 & 3.0 &  & & 600 & 3.5 & 164\\
\hline
\end{tabular}
}\label{t06}
\begin{tabnote}
\hangindent6pt\noindent
\hbox to6pt{\footnotemark[$*$]\hss}\unskip%
We assume $50\times50$ kpc since the beam size of \timeform{47''} corresponds to $\sim$52 kpc.
\end{tabnote}
\end{table*}

Hereafter, although the effect of the missing flux could be significant, we do not exclude the data above $\sim 3$~GHz in our analyses. This aims at studying the case that the variation of the fractional polarization at $\sim 3$~GHz is real. Again, we suggest to perform single dish observations at 3--3.5~GHz in the future, to clarify the effect of the missing flux on the fractional polarization.

Figure~\ref{f08} implies that the histogram of RM follows the Gaussian distribution, which is not inconsistent to consider that the beam depolarization is induced by random (turbulent) magnetic fields \citep{1982ApJ...252...81L}. As already described in Section \ref{s5ss5} and Figure~\ref{f05a}, the Burn's law with a single depolarization component is hard to reproduce the observed fractional polarization of the radio relic. Therefore, we consider models with multiple depolarization components along the line of sight toward the radio relic. A weakness of adopting the Burn's law is that we cannot extract the information of magnetic fields. In order to understand the nature of depolarization as well as magnetic fields, we hence carried out simulations of depolarization using simple random-field models. 

We calculated the polarized emission which pass the depolarization components in the model. The components consist of grid, and each grid at the three dimensional coordinate $(\mathrm{X, Y, Z})=(N_\mathrm{X}~N_\mathrm{Y}~N_\mathrm{Z}$) has single-scale random magnetic fields with uniform strength. The electron density is uniform in the components. The polarized intensity in each cylinder at $(x,y)$ is given by
\begin{equation}
P(x, y)=\int p_0 \epsilon e^{2i\phi(x,y)} dz \label{s5ss6f1}
\end{equation}
where $p_0$ is the intrinsic fractional polarization and $\epsilon$ is the synchrotron emissivity at a depth along the line of sight  \citep{1966MNRAS.133...67B, 1966ARA&A...4..245G, 1998MNRAS.299..189S}. The polarization angle $\phi(x,y)$ is given by
\begin{equation}
\phi(x, y)=\phi_0+812\sum_{z=1}^{N_\mathrm{Z}} n_\mathrm{e}B_{\parallel} \Delta l \lambda ^2 \label{s5ss6f2}
\end{equation}
where $\phi_0$ is the intrinsic polarization angle in radians, $n_\mathrm{e}$ is thermal electron density in $\mathrm{cm}^{-3}$, $B_{\parallel}$ is magnetic field strengths parallel to the line of sight in $\mu \mathrm{G}$, $\Delta l$ is the size of cells in kpc and $\lambda$ is observation wavelength in m. We obtain the polarized intensity $P$ through the $N_\mathrm{X}\times N_\mathrm{Y}$ cylinders as
\begin{equation}
P=\sum_{x=1}^{N_\mathrm{X}}\sum_{y=1}^{N_\mathrm{Y}}P(x,y). \label{s5ss6f3}
\end{equation}

Our depolarization models include the following parameters: magnetic field strengths of a cell $B$, electron density $n_\mathrm{e}$, the size of the cells $\Delta l$, the numbers of cells in the directions of the X, Y and Z axes ($N_\mathrm{X},~N_\mathrm{Y},~N_\mathrm{Z}$), and the intensity of the polarized source or emitting depolarization source. Adopting suitable parameters, we can control the optimum frequency \citep{2011MNRAS.418.2336A} and the fractional polarization.

\begin{figure}[tp]
\begin{center}
\FigureFile(80mm,50mm){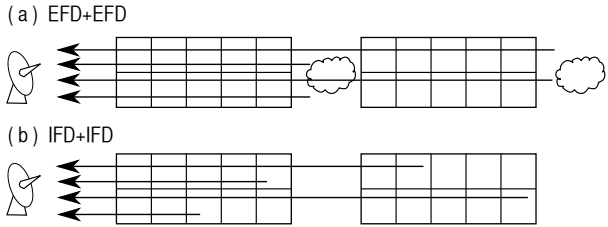}
\end{center}
\caption{(a) The depolarization model of EFD+EFD, which has non-emitting two depolarization components with two polarized sources. (b) The depolarization model of IFD+IFD, which has emitting two depolarization components. In each model, the cells constituting the depolarization components have single-scale random magnetic fields with uniform strength and uniform electron densities.}  \label{f11}
\end{figure}

We consider two components, EFD and IFD, and develop two two-component depolarization models, EFD+EFD and IFD+IFD (Figure \ref{f11}). The order of the components along the line of sight from the observer is:
\begin{itemize}
\item In the EFD+EFD model, we allocate the components in the order of a depolarization component, a polarized source, a depolarization component, and a polarized source. 
\item In the IFD+IFD model, we allocate the components in the order of an emitting depolarization component, and another emitting depolarization component.
\end{itemize}

\begin{figure}[tp]
\begin{center}
\FigureFile(80mm,50mm){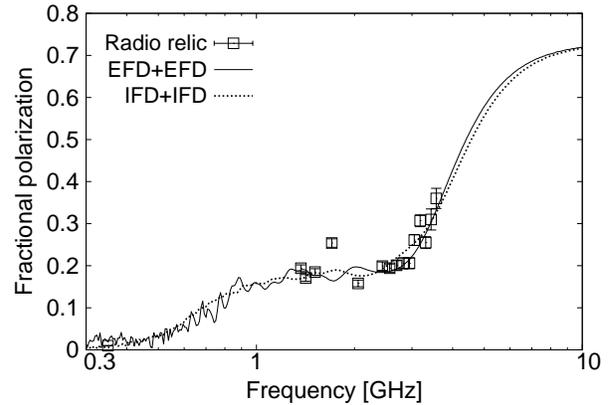}
\end{center}
\caption{The fractional polarization spectra of the depolarization models. The solid and dotted lines show the EFD+EFD model and IFD+IFD model, respectively. Open squares represent the observed fractional polarization of the radio relic (the same as Figure~\ref{f05a}).}  \label{f09}
\end{figure}

Figure~\ref{f09} shows the best-fits for the two models. Each parameter is listed in Table \ref{t06}. The EFD+EFD and IFD+IFD models can nicely reproduce the fractional polarization of the radio relic. We confirmed that we need two depolarization components to produce the step-like variations of the fractional polarization. In addition, we found that the $\sigma _{\mathrm{RM}}$ of the foreside depolarization component viewed from the observer has to be smaller than the that of the backside depolarization component. 

We could interpret the two depolarization components of the models as follows. The foreside depolarization component viewed from the observer could be the magneto-ionic plasma in the cluster or the Galaxy. The backside depolarization component should be magneto-ionic plasma inside the radio relic, according to the fact that the fractional polarization of Sources A and B does not show the same variation as the fractional polarization of the radio relic. Otherwise, if the cluster or the Galaxy depolarizes relic's polarization, the fractional polarizations of Sources A and B should also show the step-like variations. But as seen in Figure~\ref{f05b}, the fractional polarizations of Sources A and B are nearly constant. The $\sigma _{\mathrm{RM}}$ for the radio relic is larger than that for the cluster or the Galaxy.

Although we can explain observed fractional polarization of the radio relic with our depolarization models as presented in this section, we should note that there should be more realistic model to match the observational data. For instance, \citet{2004A&A...424..429M} shows a realistic model obtained by numerical simulations, including the magnetic field strength, radial profile, and magnetic field power spectrum of cluster of galaxies, which is successfully applied to individual clusters (e.g. \cite{2006A&A...460..425G,2008A&A...483..699G,2010A&A...513A..30B,2012A&A...540A..38V}).

\subsection{Faraday Tomography}
\label{s5ss7}

We also carried out Faraday tomography to make sure the existence of multiple components toward the radio relic. We apply the so-called QU-fit in which a model is fitted with the data in Stokes $Q$ and $U$ spaces (see e.g., \cite{2014PASJ...66....5I}). We perform the QU-fit using a Markov Chain Monte Carlo (MCMC) approach, so as to explore the best-set of model parameters. As for structures of Faraday components, we consider a delta function or a Gaussian. Here, the delta function consists of the three parameters: the Faraday depth $\phi$, the amplitude, and the intrinsic polarization angle $\chi_0$, and the Gaussian function consists of the above three parameters plus the width of the Gaussian (the standard deviation of the normal distribution). We consider five models, one delta function, one Gaussian, two delta functions, two Gaussians, and one delta function plus one Gaussian, then find the best model according to the Bayesian information criterion (BIC). We also check the reduced chi-square (RCS) of the best fit for each model.

\begin{table*}[tp]
\caption{The reduced chi-square (RCS), the Bayesian information criterion (BIC), and best-fit values and 1-$\sigma$ confidence regions for model parameters in the QU-fit.}\label{t04}
\begin{center}
\begin{tabular}{lcccccc}
\hline\hline
\noalign{\smallskip}
Model & RCS & BIC & $\phi$ & Amp. & $\chi_0$ & Width \\
\noalign{\smallskip}
\hline
\noalign{\smallskip}
Delta function &  21.2 & 645.2 & $ -41.55 _ { -0.742 } ^ { 0.745 } $ & $   0.42 _ { -0.008 } ^ { 0.008 } $ & $  -0.55 _ { -0.012 } ^ { 0.012 } $ & \\
Gaussian &  21.2 & 648.6 & $ -41.53 _ { -0.775 } ^ { 0.736 } $ & $   0.42 _ { -0.008 } ^ { 0.008 } $ & $  -0.55 _ { -0.011 } ^ { 0.012 } $ & $   0.02 _ { -0.005 } ^ { 0.736 } $ \\
two Deltas &   3.0 & 110.2 & $ -40.87 _ { -0.016 } ^ { 0.626 } $ & $  21.23 _ { -1.464 } ^ { -1.118 } $ & $   0.12 _ { -0.010 } ^ { -0.003 } $ & \\
& & & $ -40.36 _ { -0.008 } ^ { 0.649 } $ & $  21.17 _ { -1.472 } ^ { -1.149 } $ & $  -1.45 _ { -0.010 } ^ { -0.003 } $ & \\
two Gaussian &   2.2 &  93.3 & $ -49.32 _ { -1.003 } ^ { 0.191 } $ & $  13.31 _ { 0.779 } ^ { 1.019 } $ & $   0.42 _ { -0.051 } ^ { -0.013 } $ & $   8.36 _ { -0.885 } ^ { 1.727 } $ \\
& & & $ -48.14 _ { -1.032 } ^ { 0.134 } $ & $  13.41 _ { 0.764 } ^ { 1.011 } $ & $  -1.15 _ { -0.050 } ^ { -0.012 } $ & $   8.70 _ { -0.733 } ^ { 1.913 } $ \\
Delta + Gaussian &   2.6 & 100.9 & $ -50.35 _ { -1.162 } ^ { 0.322 } $ & $  12.11 _ { -1.555 } ^ { -1.270 } $ & $   0.55 _ { 0.045 } ^ { 0.069 } $ & \\
& & & $ -49.38 _ { -1.066 } ^ { 0.320 } $ & $  12.28 _ { -1.536 } ^ { -1.244 } $ & $  -1.02 _ { 0.045 } ^ { 0.068 } $ & $   1.50 _ { -1.341 } ^ { 0.217 } $ \\
\noalign{\smallskip}
\hline
\end{tabular}
\end{center}
\end{table*}

\begin{figure}[tp]
\begin{center}
\FigureFile(80mm,50mm){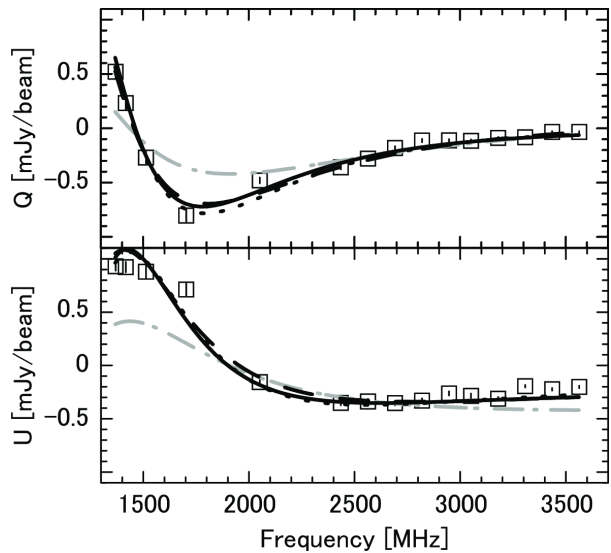}
\end{center}
\caption{
Results of the QU-fit. Marks with error bars show the spatially-averaged fractional polarization data (Figures~\ref{f05a} and \ref{f09}). Gray lines show the one-component fits (dashed: delta function, dotted: Gaussian). Black lines show the two-component fits (dashed: two delta functions, dotted: two Gaussians, solid: one delta function plus one Gaussian).
}
\label{f10}
\end{figure}

The results are shown in Figure \ref{f10} and Table \ref{t04}. We find that one-component models poorly reproduce the observed Q and U, and apparently the two-component models better fit with the data. Actually, two-component models dramatically improve BICs and RCSs (Table \ref{t04}). We do not conclude the best-model, since the two-component models show similar BICs with each other. RCSs of the two-component models are a bit far from unity. To improve the fit, data below $\sim 1500$~MHz is essential.

The fits with the two-component models commonly suggest that there are components at the Faraday depth $\phi \sim -50$~${\rm rad~m^{-2}}$. This would be the radio relic, since the depth and the thickness are respectively close to the average and the standard deviation of RM for the radio relic (Table~\ref{t03}). 

\section{Conclusions} \label{s6}

We reported new polarimetry results of Abell 2256 with JVLA at S-band (2051--3947 MHz) and X-band (8051--9947 MHz) in the C array configuration. We made images of the Stoke $I$, $Q$, and $U$, with \timeform{47''} and \timeform{15.1''} resolutions. At S-band, we detected the significant polarized emission from the radio relic, Source A, and Source B. At X-band, we detected the significant polarized emission only from Source A.

The total 2051 MHz flux of the radio relic is 286~mJy which includes the flux from the tail of Source C. The total flux is substantially smaller than $\sim 330$~mJy, an expectation from previous L-band observations with an assumption of the spectral index $\alpha =-0.81$. The estimated upper limit of the total 2051 MHz flux of the radio halo is $\sim49~\mathrm{mJy}$ assuming that the radius of the radio halo emission is $\sim\timeform{6.1'}$ and the upper limit of the flux density is $606~\mathrm{\mu Jy~beam^{-1}}$. The estimated flux is also smaller than $\sim66$ mJy, an expectation from previous L-band observations with an assumption of the spectral index $\alpha =-1.1$.

We examined the missing flux caused by the largest angular scale (LAS) of our observations. Actually, the scale of the major axis of the radio relic and halo are \timeform{1014''} and \timeform{732''} \citep{2006AJ....131.2900C} on the sky, respectively, while the LAS is \timeform{970''} at 1.5 GHz or \timeform{490''} at 3.0 GHz in the JVLA C array configuration.

We obtained RMs of the radio relic, Source A, and Source B. The mean and standard deviation of RM are $\langle \mathrm{RM_{relic}} \rangle =-34.5~ \mathrm{rad ~ m}^{-2}$ and $\sigma_{\mathrm{RM_{relic}}}=6.2~ \mathrm{rad ~ m}^{-2}$ in the radio relic, $\langle \mathrm{RM_A} \rangle=-24.9~ \mathrm{rad ~ m}^{-2}$ and $\sigma_{\mathrm{RM_A}}=65.5~ \mathrm{rad ~ m}^{-2}$ in the source A, and $\langle \mathrm{RM_B} \rangle=-34.1~ \mathrm{rad ~ m}^{-2}$ and $\sigma_{\mathrm{RM_\mathrm{B}}}=10.5~ \mathrm{rad ~ m}^{-2}$ in the source B.

We calculated the magnetic field strengths in the radio relic and the intracluster space in Abell 2256. For the radio relic, we calculated the magnetic field strengths using the revised equipartition formula, and the fractional polarization. The total magnetic field strengths is $\sim5.0~\mu\mathrm{G}$ with $l=25~\mathrm{kpc}$ and $\sim1.8~\mu\mathrm{G}$ with $l=1125\mathrm{kpc}$, the uniform magnetic field strength is $\sim3.7~\mathrm{\mu G}$ with $l=25~\mathrm{kpc}$ and $\sim1.3~\mathrm{\mu G}$ with $l=1125~\mathrm{kpc}$, and the random magnetic field strength is $\sim3.4~\mathrm{\mu G}$ with $l=25~\mathrm{kpc}$ and $\sim1.2~\mathrm{\mu G}$ with $l=1125~\mathrm{kpc}$. For the intracluster space, we calculated the magnetic field strengths using $\sigma_{\mathrm{RM}}$. The magnetic-field strengths along the line of sight toward Source A is $B\sim 1.26 ~\mu \mathrm{G}$ with $\Lambda_\mathrm{B}=5~ \mathrm{kpc}$ and $B\sim 0.63~\mu \mathrm{G}$ with $\Lambda_\mathrm{B}=20~\mathrm{kpc}$. The magnetic field strengths along the line of sight toward Source B is $B\sim 0.21 ~\mu \mathrm{G}$ with $\Lambda_\mathrm{B}=5~\mathrm{kpc}$ and $B\sim 0.11 ~\mu \mathrm{G}$ with $\Lambda_\mathrm{B}=20 ~\mathrm{kpc}$.

We inferred that the shift of the mean of the RM from $0~\mathrm{rad~m^{-2}}$ to about $-30~\mathrm{rad~m^{-2}}$ toward the Abell 2256 field is due to the Galactic contribution since the averaged RM values of 28 polarized sources within \timeform{6D} around Abell 2256 is broadly consistent with the means of RM for the radio relic, Source A, and Source B.

We found that the fractional polarization of the radio relic remains about 20~\% between 1.3--3~GHz and increases above 3 GHz. The Burn's law with a single depolarization component cannot reproduce the observed step-like fractional polarization spectrum of the radio relic. This may be due to the missing flux and/or depolarization.

Our simulations of depolarization, which allow us to know three dimensional position information of the magneto-ionic plasma from the fractional polarization, indicated that two-component depolarization models can explain the step-like variations of the fractional polarization. Furthermore, we found that the standard deviation of RM for the foreside component viewed from the observer should be smaller than that of the backside component. The existence of two components was also suggested from Faraday tomography.

\bigskip
This work was supported by Japan Society for Promotion of Science KAKENHI Grant Numbers 26800104 (HN), 26400218 (MT), 15K17614 and 15H03639 (TA). The authors would like to thank S. Ideguchi for providing a QU-fit code. 



\begin{thebibliography}{}
\bibitem[Arshakian \& Beck(2011)]{2011MNRAS.418.2336A} Arshakian, T.~G., \& Beck, R.\ 2011, \mnras, 418, 2336 
\bibitem[Beck(1982)]{1982A&A...106..121B} Beck, R.\ 1982, \aap, 106, 121 
\bibitem[Beck \& Krause(2005)]{2005AN....326..414B} Beck, R., \& Krause, M.\ 2005, Astronomische Nachrichten, 326, 414 
\bibitem[Berrington et al.(2002)]{2002AJ....123.2261B} Berrington, R.~C., Lugger, P.~M., \& Cohn, H.~N.\ 2002, \aj, 123, 2261 
\bibitem[Bonafede et al.(2010)]{2010A&A...513A..30B} Bonafede, A., Feretti, L., Murgia, M., et al.\ 2010, \aap, 513, A30
\bibitem[Brentjens(2008)]{2008A&A...489...69B} Brentjens, M.~A.\ 2008, \aap, 489, 69 
\bibitem[Bridle \& Fomalont(1976)]{1976A&A....52..107B} Bridle, A.~H., \& Fomalont, E.~B.\ 1976, \aap, 52, 107 
\bibitem[Bridle et al.(1979)]{1979A&A....80..201B} Bridle, A.~H., Fomalont, E.~B., Miley, G.~K., \& Valentijn, E.~A.\ 1979, \aap, 80, 201 
\bibitem[Briel et al.(1991)]{1991A&A...246L..10B} Briel, U.~G., Henry, J.~P., Schwarz, R.~A., et al.\ 1991, \aap, 246, L10 
\bibitem[Briel \& Henry(1994)]{1994Natur.372..439B} Briel, U.~G., \& Henry, J.~P.\ 1994, \nat, 372, 439 
\bibitem[Brunetti et al.(2001)]{2001MNRAS.320..365B} Brunetti, G., Setti, G., Feretti, L., \& Giovannini, G.\ 2001, \mnras, 320, 365
\bibitem[Brunetti et al.(2004)]{2004MNRAS.350.1174B} Brunetti, G., Blasi, P., Cassano, R., \& Gabici, S.\ 2004, \mnras, 350, 1174
\bibitem[Buote(2001)]{2001ApJ...553L..15B} Buote, D.~A.\ 2001, \apjl, 553, L15 
\bibitem[Burn(1966)]{1966MNRAS.133...67B} Burn, B.~J.\ 1966, \mnras, 133, 67
\bibitem[Cassano \& Brunetti(2005)]{2005MNRAS.357.1313C} Cassano, R., \& Brunetti, G.\ 2005, \mnras, 357, 1313 
\bibitem[Chen et al.(2007)]{2007A&A...466..805C} Chen, Y., Reiprich, T.~H., B{\"o}hringer, H., Ikebe, Y., \& Zhang, Y.-Y.\ 2007, \aap, 466, 805 
\bibitem[Cho \& Ryu(2009)]{2009ApJ...705L..90C} Cho, J., \& Ryu, D.\ 2009, \apjl, 705, L90
\bibitem[Clarke \& Ensslin(2006)]{2006AJ....131.2900C} Clarke, T.E., \& Ensslin, T.A.\ 2006, \aj, 131, 2900
\bibitem[Donnert et al.(2013)]{2013MNRAS.429.3564D} Donnert, J., Dolag, K., Brunetti, G., \& Cassano, R.\ 2013, \mnras, 429, 3564
\bibitem[Ebeling et al.(1998)]{1998MNRAS.301..881E} Ebeling, H., Edge, A.~C., Bohringer, H., et al.\ 1998, \mnras, 301, 881
\bibitem[Felten(1996)]{1996ASPCP...88...271} Felten, J. E. 1996, Clusters, Lensing, and the Future of the Universe, 88, 271
\bibitem[Feretti et al.(1995)]{1995A&A...302..680F} Feretti, L., Dallacasa, D., Giovannini, G., \& Tagliani, A.\ 1995, \aap, 302, 680 
\bibitem[Feretti et al.(2012)]{2012A&ARv..20...54F} Feretti, L., Giovannini, G., Govoni, F., \& Murgia, M.\ 2012, \aapr, 20, 54 
\bibitem[Fujita et al.(2003)]{2003ApJ...584..190F} Fujita, Y., Takizawa, M., \& Sarazin, C.~L.\ 2003, \apj, 584, 190
\bibitem[Gardner \& Whiteoak(1966)]{1966ARA&A...4..245G} Gardner, F.~F., \& Whiteoak, J.~B.\ 1966, \araa, 4, 245 
\bibitem[Govoni et al.(2006)]{2006A&A...460..425G} Govoni, F., Murgia, M., Feretti, L., et al.\ 2006, \aap, 460, 425 
\bibitem[Govoni et al.(2010)]{2010A&A...522A.105G} Govoni, F., Dolag, K., Murgia, M., et al.\ 2010, \aap, 522, AA105 
\bibitem[Guidetti et al.(2008)]{2008A&A...483..699G} Guidetti, D., Murgia, M., Govoni, F., et al.\ 2008, \aap, 483, 699 
\bibitem[Haslam et al.(1978)]{1978A&AS...31...99H} Haslam, C.~G.~T., Kronberg, P.~P., Waldthausen, H., Wielebinski, R., \& Schallwich, D.\ 1978, \aaps, 31, 99 
\bibitem[Ideguchi et al.(2014)]{2014PASJ...66....5I} Ideguchi, S., Takahashi, K., Akahori, T., Kumazaki, K., \& Ryu, D.\ 2014, \pasj, 66, 5 
\bibitem[Kale \& Dwarakanath(2010)]{2010ApJ...718..939K} Kale, R., \& Dwarakanath, K.~S.\ 2010, \apj, 718, 939 
\bibitem[Lawler \& Dennison(1982)]{1982ApJ...252...81L} Lawler, J.~M., \& Dennison, B.\ 1982, \apj, 252, 81 
\bibitem[Miller et al.(2003)]{2003AJ....125.2393M} Miller, N.~A., Owen, F.~N., \& Hill, J.~M.\ 2003, \aj, 125, 2393
\bibitem[Murgia et al.(2004)]{2004A&A...424..429M} Murgia, M., Govoni, F., Feretti, L., et al.\ 2004, \aap, 424, 429 
\bibitem[Ohno et al.(2002)]{2002ApJ...577..658O} Ohno, H., Takizawa, M., \& Shibata, S.\ 2002, \apj, 577, 658
\bibitem[Owen(1975)]{1975AJ.....80..263O} Owen, F.~N.\ 1975, \aj, 80, 263 
\bibitem[Owen et al.(2014)]{2014ApJ...794...24O} Owen, F.~N., Rudnick, L., Eilek, J., et al.\ 2014, \apj, 794, 24
\bibitem[Petrosian(2001)]{2001ApJ...557..560P} Petrosian, V.\ 2001, \apj, 557, 560
\bibitem[Rottgering et al.(1994)]{1994ApJ...436..654R} Rottgering, H., Snellen, I., Miley, G., et al.\ 1994, \apj, 436, 654
\bibitem[Ryu et al.(2008)]{2008Sci...320..909R} Ryu, D., Kang, H., Cho, J., \& Das, S.\ 2008, Science, 320, 909 
\bibitem[Segalovitz et al.(1976)]{1976Natur.264..222S} Segalovitz, A., Shane, W.~W., \& de Bruyn, A.~G.\ 1976, \nat, 264, 222
\bibitem[Sokoloff et al.(1998)]{1998MNRAS.299..189S} Sokoloff, D.~D., Bykov, A.~A., Shukurov, A., et al.\ 1998, \mnras, 299, 189 
\bibitem[Sun et al.(2002)]{2002ApJ...565..867S} Sun, M., Murray, S.~S., Markevitch, M., \& Vikhlinin, A.\ 2002, \apj, 565, 867
\bibitem[Takizawa \& Naito(2000)]{2000ApJ...535..586T} Takizawa, M., \& Naito, T.\ 2000, \apj, 535, 586
\bibitem[Tamura et al.(2011)]{2011PASJ...63S1009T} Tamura, T., Hayashida, K., Ueda, S., \& Nagai, M.\ 2011, \pasj, 63, 1009
\bibitem[Taylor et al.(2009)]{tss09} Taylor, A. R., Stil, J. M., \& Sunstrum, C. 2009, \apj, 702, 1230
\bibitem[Trasatti et al.(2015)]{2015A&A...575A..45T} Trasatti, M., Akamatsu, H., Lovisari, L., et al.\ 2015, \aap, 575, A45
\bibitem[Tribble(1991)]{1991MNRAS.250..726T} Tribble, P.~C.\ 1991, \mnras, 250, 726
\bibitem[Vacca et al.(2012)]{2012A&A...540A..38V} Vacca, V., Murgia, M., Govoni, F., et al.\ 2012, \aap, 540, A38
\bibitem[van Weeren et al.(2009)]{2009A&A...508.1269V} van Weeren, R.~J., Intema, H.~T., Oonk, J.~B.~R., R{\"o}ttgering, H.~J.~A., \& Clarke, T.~E.\ 2009, \aap, 508, 1269
\bibitem[van Weeren et al.(2012)]{2012A&A...543A..43V} van Weeren, R.~J., R{\"o}ttgering, H.~J.~A., Rafferty, D.~A., et al.\ 2012, \aap, 543, A43 
\bibitem[Vazza et al.(2009)]{2009MNRAS.395.1333V} Vazza, F., Brunetti, G., \& Gheller, C.\ 2009, \mnras, 395, 1333
\bibitem[Xu et al.(2009)]{2009ApJ...698L..14X} Xu, H., Li, H., Collins, D.~C., Li, S., \& Norman, M.~L.\ 2009, \apjl, 698, L14
\bibitem[Xu et al.(2010)]{2010ApJ...725.2152X} Xu, H., Li, H., Collins, D.~C., Li, S., \& Norman, M.~L.\ 2010, \apj, 725, 2152
\end{thebibliography}
\end{document}